\documentclass[preprint,12pt,authoryear]{elsarticle}

\usepackage{amssymb}
\usepackage{color}

\journal{Icarus}

\begin{document}

\begin{frontmatter}



\address[arc]{NASA Ames Research Center, Moffett Field, CA 94035}
\address[uconn]{Department of Physics, University of Connecticut, Storrs, CT 06268}
\address[itamp]{Institute for Theoretical Atomic and Molecular Physics, Harvard-Smithsonian Center for Astrophysics, Cambridge, MA 02138}

\title{Non-thermal production and escape of OH from the upper atmosphere of Mars}


\author[arc]{M.~Gacesa}
\ead{marko.gacesa@nasa.gov}
\author[uconn]{N.~Lewkow}
\ead{nlewkow@gmail.com}
\author[itamp,uconn]{V.~Kharchenko}
\ead{kharchenko@cfa.harvard.edu}

\address{}

\begin{abstract} 
We present a theoretical analysis of formation and kinetics of hot OH molecules in the upper atmosphere of Mars produced in reactions of thermal molecular hydrogen and energetic oxygen atoms. Two major sources of energetic O considered are the photochemical production, via dissociative recombination of O$_{2}^{+}$ ions, and energizing collisions with fast atoms produced by the precipitating Solar Wind (SW) ions, mostly H$^+$ and He$^{2+}$, and energetic neutral atoms (ENAs) originating in the charge-exchange collisions between the SW ions and atmospheric gases. Energizing collisions of O  with  atmospheric secondary hot atoms, induced by precipitating  SW ions and ENAs, are also included in our consideration. 
The non-thermal reaction O + H$_2(v,j) \rightarrow$ H + OH$(v',j')$ is described using recent quantum-mechanical state-to-state cross sections, which allow us to predict non-equilibrium distributions of excited rotational and vibrational states $(v',j')$ of OH and expected emission spectra. A fraction of produced translationally hot OH is sufficiently energetic to overcome Mars' gravitational potential and escape into space, contributing to the hot corona. We estimate the total escape flux from dayside of Mars for low solar activity conditions at about $5\times10^{22}$ s$^{-1}$, or about 0.1\% of the total escape rate of atomic O and H.
The described non-thermal OH production mechanism is general and expected to contribute to the evolution of atmospheres of the planets, satellites, and exoplanets with similar atmospheric compositions.
\end{abstract}


\end{frontmatter}


\section{Introduction}

The escape of volatile atmospheric species to space is important for understanding the evolution of Mars' atmosphere and climate as it transitioned from the conditions that supported liquid water into the cold, dry, low-pressure climate that we witness today \citep{1972Sci...177..986M,2008SSRv..139..355J,2013SSRv..174..113L,2015SSRv..195..357L}. While it is well established that the evaporation of the martian atmosphere is driven by the interaction with the solar radiation and interplanetary plasma, with the absence of an intrinsic planetary magnetic field and Mars' lower mass accelerating the process \citep{2004P&SS...52.1039C,2008SSRv..139..355J}, detailed physical mechanisms and their mutual interactions are still not fully understood and 3D global atmospheric models cannot simultaneously explain all observed effects \citep{2015SSRv..195..357L,2015GeoRL..42.9015L,2015JGRE..120.1880L}. 
Attempting to resolve the remaining unanswered questions and shed light on water inventory in the early history of Mars is the main scientific objective of the ongoing NASA's Mars Atmosphere and Volatile Evolution (MAVEN) mission \citep{2015SSRv..195....3J,2015SSRv..195..423B}.

At the present time, the atmospheric escape from Mars is comprised of thermal (Jeans) escape and various non-thermal mechanisms, including photo-chemical escape of neutrals \citep{2004P&SS...52.1039C,2008SSRv..139..355J,2013SSRv..174..113L,2015JGRE..120.1880L} and escape of ions governed by the interplay of the solar wind with the induced martian magnetosphere and crustal magnetic fields \citep{acuna1999global,2004SSRv..111...33N,2015JGRA..120.7857D,2015GeoRL..42.8870R}. Major escaping species include atomic hydrogen, oxygen, and carbon, of which the first two directly affect the estimates of water abundance on primordial Mars \citep{2013SSRv..174..113L}. 

A major photochemical process responsible for escape of neutrals heavier than hydrogen is dissociative recombination (DR) of O$_{2}^{+}$, which serves as a major source of hot O atoms that either directly escape to space or form a hot oxygen corona \citep{1988Icar...76..135I,1993GeoRL..20.1747F,2005SoSyR..39...22K,2015JGRE..120.1880L,2015GeoRL..42.9009D}. 
The nascent suprathermal O atoms can collide with thermal background gases in the upper atmosphere and transfer sufficient kinetic energy to eject them to space. This non-thermal escape mechanism of light elements, also known as a  collisional ejection, was studied for He atoms \citep{2011GeoRL..3802203B}, and H$_2$ and HD molecules \citep{2012GeoRL..3910203G}, which were found to produce significant fluxes and possibly affect the H/D ratio in Mars' upper atmosphere by 5-10\%.

One of the major goals of our research, reported in this article, is to develop a consistent model of production of non-thermal atoms and molecules in the Martian atmosphere and describe non-equilibrium atmospheric reactions caused by hot particles. 
In this study, we explore reactive collisions of hot O atoms with H$_2$ molecules, leading to formation of rotationally-vibrationally (RV) excited OH molecules in Mars' upper atmosphere. The description of the reaction is based on quantum-mechanical state-to-state reactive cross sections at high temperatures \citep{2014JChPh.141p4324G}, while kinetic theory is used to calculate the energy transfer to translational and internal degrees of freedom of the products. 
We use a 1D model of the martian atmosphere to estimate altitude profiles of the total formation and escape rates of OH molecules and non-thermal RV distributions for a selected orbital geometry and solar activity. In addition to the DR, secondary hot O atoms energized in collisions with energetic neutral atoms (ENAs) \citep{2014ApJ...790...98L} are considered as an efficient source of hot O atoms in our model.

\section{Methods}

\subsection{Hot atom sources}

We consider dissociative recombination (DR) of oxygen molecular ions with electrons, O$_{2}^{+} + e \rightarrow$ O + O + $\Delta E$, as the main source of hot O atoms. This process is widely accepted as the dominant source of O in the martian ionosphere in the current epoch \citep{1988Icar...76..135I,1993GeoRL..20.1747F,2005SoSyR..39...22K}. The DR of O$_{2}^{+}$ is an exothermic process that can proceed via five dissociation pathways and produce energetic O($^3$P), O($^1$D), and O($^1$S) with excess kinetic energy between 0.84 eV and 6.99 eV per a pair \citep{1988P&SS...36...47G,2009Icar..204..527F}. 
Since the state-to-state cross sections are available only for the reaction of O($^3$P) + H$_2$, we assume that all reactive collisions involve hot O($^3$P). This approximation is further justified by the fact that the lifetime of energetic metastable O($^1$D) atoms is short before they relax into their ground state via spontaneous emission or collisional quenching with atmospheric gases \citep{2005JGRA..11012305K}. 

To evaluate the contribution of the photo-chemical production mechanism of energetic O atoms, we followed the approach of \cite{2011GeoRL..3802203B} to construct the energy distribution and rate of production of hot O atoms, $f_\mathrm{DR}(E_i, h)$, as a function of the altitude $h$ and the initial energy $E_i$ of the O atoms dissociated via $i$-th channel. Four significant dissociation channels were considered with branching ratios and exothermicities given by \cite{1988P&SS...36...47G} and \cite{2009Icar..204..527F}. We constructed $f_\mathrm{DR}(E_i, h)$ distributions of nascent O atoms below 400 km for conditions of low solar activity \citep{2009Icar..204..527F} and adopted the atmospheric model by \cite{2010Icar..207..638K} with extrapolated hot O source functions from \cite{2009Icar..204..527F}.

\begin{figure}[t]
  \noindent\includegraphics[width=0.95\columnwidth]{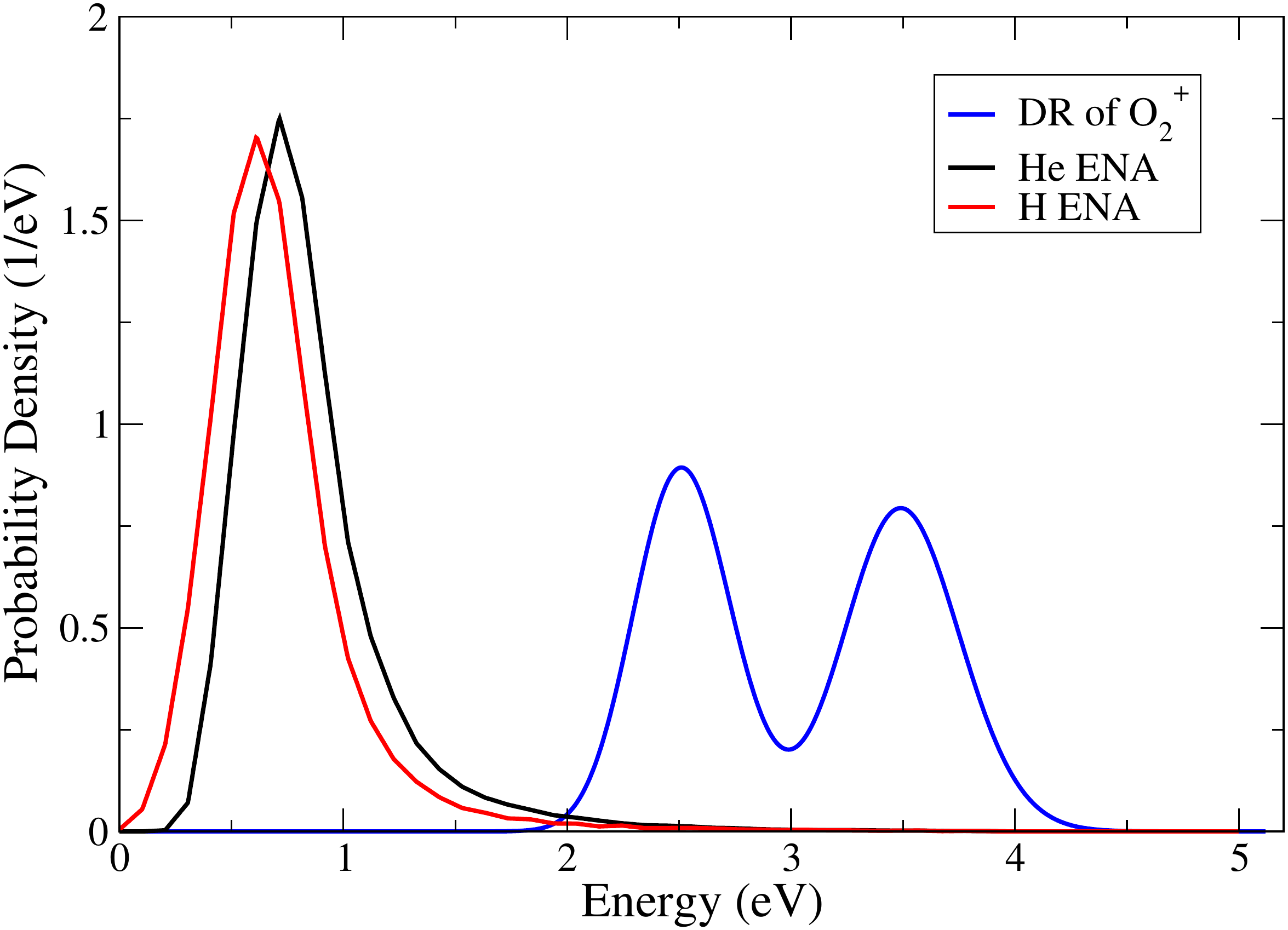}  
  \caption{Energy distributions of hot O($^3P$) produced by DR of O$_2^+$ and in collisions with energetic neutral H and He atoms. The distributions are normalized to unity and calculated at 200 km.}
\label{fig1}
\end{figure}

As a secondary source of hot O atoms we considered elastic collisions of precipitating ENAs, namely energetic H and He, with thermal atmospheric oxygen, leading to production of secondary hot O atoms (SHOAs) \citep{2005SoSyR..39...22K,2013JGRA..118.7635W,2014ApJ...790...98L}. Incoming ENAs decelerate in the martian atmosphere by transferring a fraction of its kinetic energy to thermal atmospheric gases typically in thousands of collisions dominated by small scattering angles. The SHOAs are produced mainly at altitudes lower than 250 km, where the atmospheric density sharply increases, and are predicted to have kinetic energies up to 4 eV \citep{2014ApJ...790...98L}. We carried out a 3D Monte Carlo simulation of thermalization of ENAs in the martian atmosphere for average solar wind activity and constructed altitude-dependent SHOA source function $f_\mathrm{ENA}(h)$. A distinct feature of our MC approach are energy transfer cross sections given as functions of both collision velocity and angle, resulting in strongly forward-peaked anisotropic distributions. The simulation is described in detail in \cite{2014ApJ...790...98L}. In this work we adopted similar physical parameters, including the ensemble sizes, as in the article. 

\begin{figure}[t]
  \noindent\includegraphics[width=0.95\columnwidth]{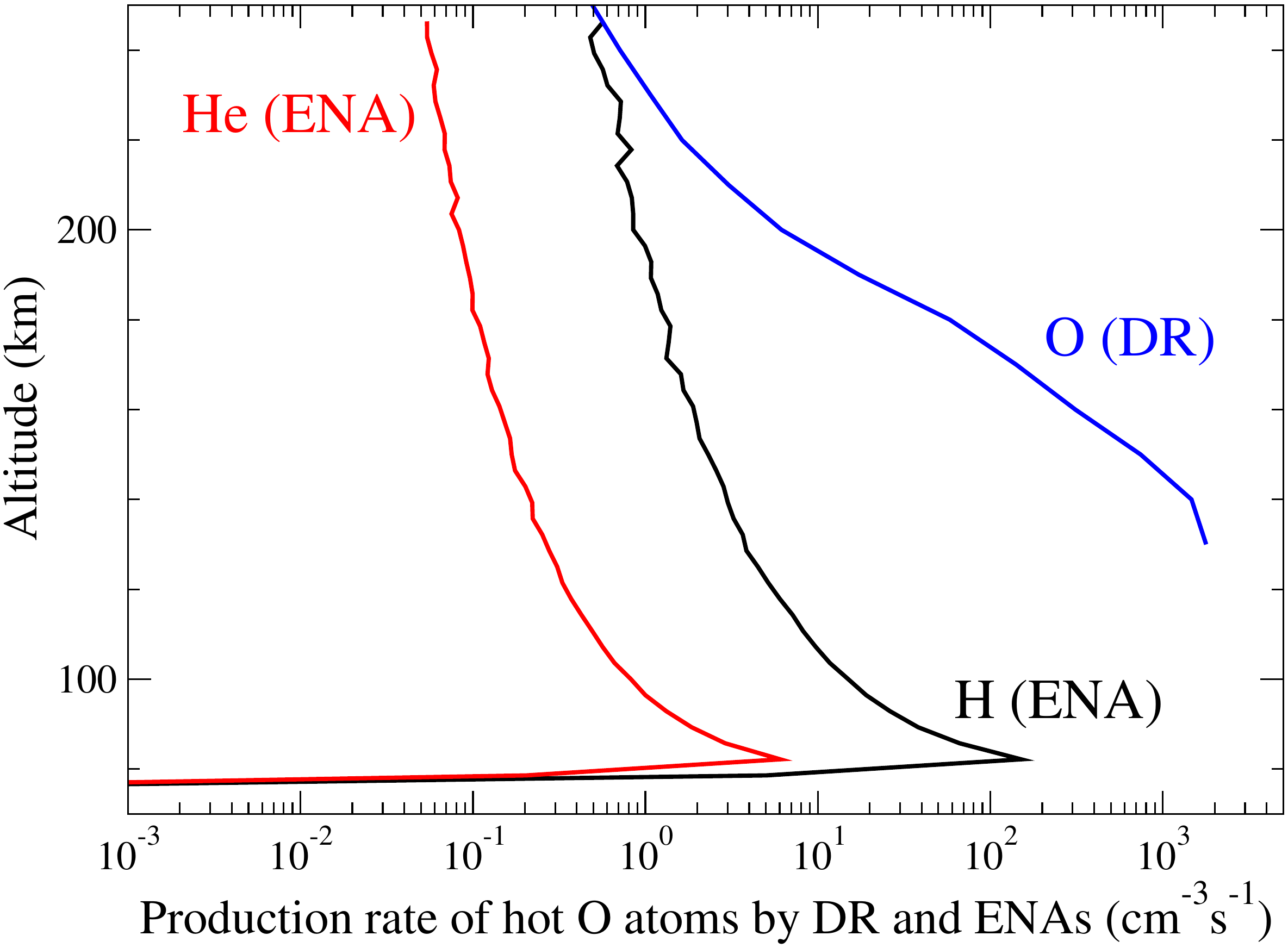}  
  \caption{Altitude profile of the rate of production of secondary hot O($^3P$) atoms by DR of O$_2^+$ and collisions of precipitating H and He ENAs with atmospheric O. Mean solar activity is assumed.}
\label{fig2}
\end{figure}

The energy distributions (normalized to unity) of hot O and its volume production rates as a function of altitude are given for DR and ENA production mechanisms in Fig. \ref{fig1} and Fig. \ref{fig2}, respectively. 
The total source function $f(E_i,h)$, given in the units of volume production rate, was constructed as a sum of distributions produced by considering both non-thermal processes, $f(E_i,h) = f_\mathrm{DR}(E_i,h) + f_\mathrm{ENA}(E_i,h)$. For the altitudes higher than 250 km and lower than 130 km (in case of DR of O$_{2}^{+}$), the production rate distributions were smoothly extrapolated with descending exponential functions to avoid unphysical discontinuities. The final results were not found to be depend much on the exact functional forms due to the fact that the production rates in the extrapolated regions are very small. The energy distributions given in Fig. \ref{fig1} were calculated for the altitude of 200 km and do not change significantly for the altitudes at which majority of thermalization processes take place \citep{2009JGRA..114.7101Z,2009Icar..204..527F}.

\subsection{Cross sections and kinetics of non-thermal OH production}

We use quantum-mechanical state-to-state reactive cross sections and differential cross sections (DCSs) \citep{2014JChPh.141p4324G} to model the chemical reaction O($^3$P)+H$_2(v,j) \rightarrow$ H + OH$(v',j')$, where $(v,j)$ and $(v',j')$ are the initial and final RV levels of H$_2$ and OH, respectively. The cross sections of \cite{2012GeoRL..3910203G} are adopted for elastic scattering. Both sets of cross sections were constructed in an extended collision energy range for purposes of modeling non-thermal atmospheric processes and high-temperature combustion.

Momentum transfer reactive cross sections are evaluated according to \citep{1978JChPh..68.1585P}
\begin{equation}
  \sigma_{vj,v'j'}^{\mathrm{mt}}(E) = 2\pi \int_{0}^{\infty} Q_{vj,v'j'}(E,\theta) \left( 1-\cos \theta \sqrt{ \frac{E'}{E} } \right) \sin \theta d\theta ,
  \label{eq:mtcs}
\end{equation}
where $Q_{vj,v'j'}(E,\theta)$ are reactive differential cross sections (DCSs) and $E' = E - (E_{v'j'}-E_{vj})$ is the final energy of the product OH. Here, the internal RV energies of H$_2(v,j)$ and OH($v',j'$) are given by $E_{vj}$ and $E_{v'j'}$, respectively. The RV energies $E_{vj}$ and $E_{v'j'}$ were found using a mapped Fourier Grid Method \citep{1999JChPh.110.9865K} in the asymptotic limits (corresponding to diatomic H$_2$ and OH) of the potential energy surfaces used in \cite{2014JChPh.141p4324G}.

The kinetic energy transfer rate from hot O atoms to the product OH$(v',j')$ is determined using kinetic theory with quantized internal molecular degrees of freedom and anisotropic cross sections as \citep{1982itam.book.....J}
\begin{equation}
  T_{v' j'} = \frac{m_\mathrm{O} \, m_{\mathrm{OH}}} {(m_\mathrm{O} + m_{\mathrm{OH}})^2} 
      \left( 1 + \gamma_{v' j'} - 2\sqrt{\gamma_{v' j'}} \cos \theta \right) E,
\label{eq1}
\end{equation}
where 
$m_\mathrm{O}$ and $m_\mathrm{{OH}}$ are masses of O and OH, respectively, $E$ is the collision energy in the laboratory frame (LF), $\gamma_{v' j'} = \epsilon_{v' j'} / \epsilon$, is the ratio of $\epsilon$ and $\epsilon_{v' j'}$, the translational kinetic energies before and after the reaction in the center-of-mass (CM) frame, respectively, and $\theta$ is the scattering angle in the CM frame. 

\begin{figure}[t]
  \noindent \includegraphics[width=0.95\columnwidth]{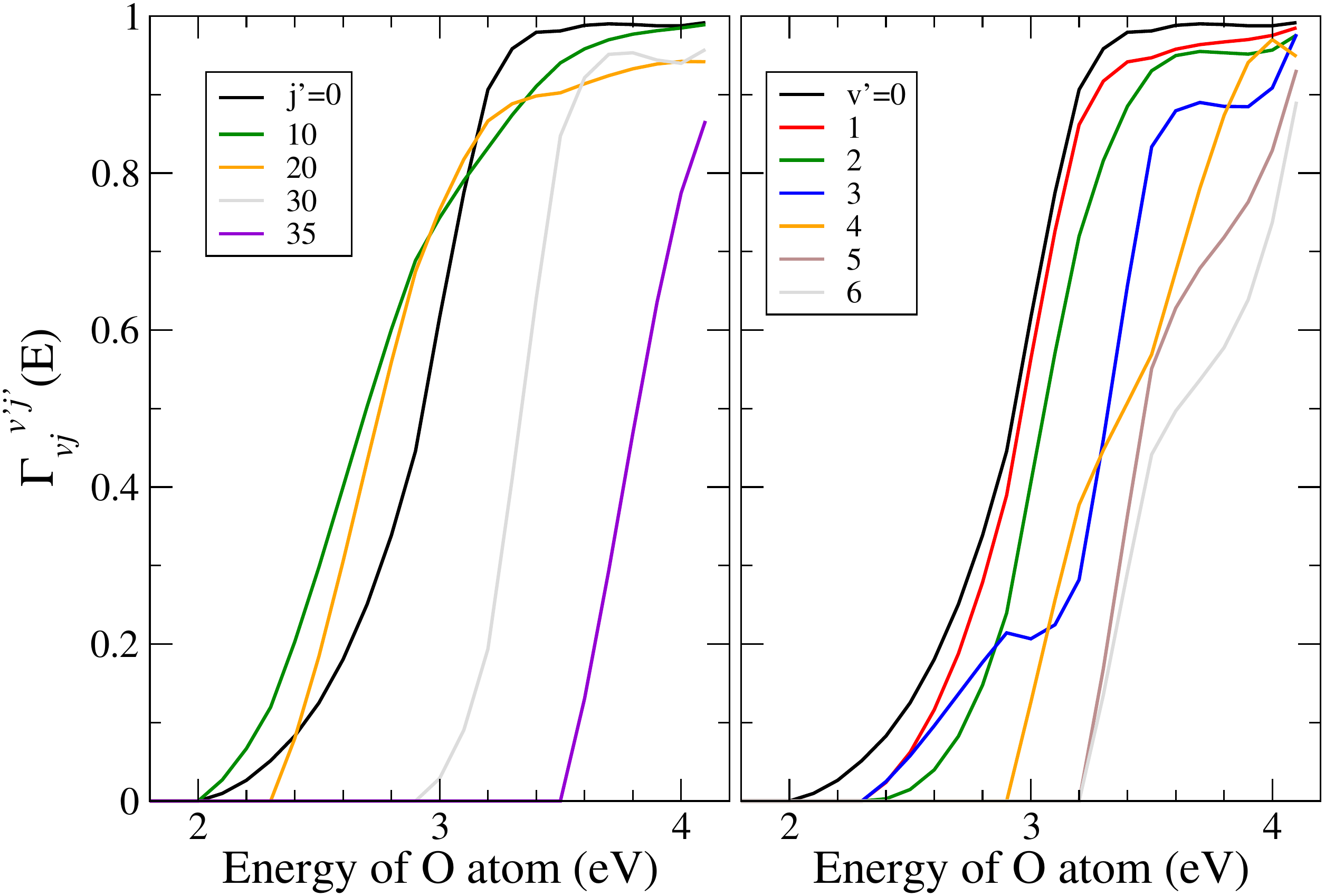}  
  \caption{Fraction $\Gamma_{vj}^{v'j'}(E)$ of the OH$(v',j')$ molecules produced in the reaction O+H$_2(v=0, j=0)$ having sufficient kinetic energy to escape from Mars given as a function of O atom kinetic energy. Selected rotationally  (left panel) and vibrationally (right panel) excited states of OH are shown.}
  \label{fig_gammas}
\end{figure}

Following the reaction, the fraction of OH molecules sufficiently energetic to escape is given by 
\begin{equation}
  \Gamma_{vj}^{v'j'}(E) = \frac{\int_{\theta_{\mathrm{min}}}^{\pi} Q_{vj,v'j'}(E,\theta) 
                 \left(1-\cos \theta \right) \sin \theta d \theta }
             {\int_{0}^{\pi} Q_{vj,v'j'}(E,\theta) \left(1-\cos \theta \right) \sin \theta d \theta } ~,
  \label{eq2}
\end{equation}
where the critical angle $\theta_{\mathrm{min}}$ was determined by solving Eq. (\ref{eq1}) for the angle, while requiring that the translational part of the transferred kinetic energy, $T_{v',j'}$ is equal to the escape energy of OH from Mars, $E^\mathrm{OH}_{\mathrm{esc}} = 2.08$ eV. The resulting fractions $\Gamma_{vj}^{v'j'}(E)$ for the vibrational states up to $v_\mathrm{max} = 6$ and $j_\mathrm{max} = 36$ were calculated and illustrated in Fig. \ref{fig_gammas} for selected values of $(v',j')$ states.
According to the kinetic model, the OH molecules could be produced in the reaction in higher RV levels but they would not receive sufficient kinetic energy to be able to escape.

\subsection{1D transport model}

To determine the transport of non-thermal atoms and molecules, including altitude profiles of OH, we use a 1D model built upon analogous assumptions used in the study of escape of neutral He atoms and H$_2$ molecules from Mars \citep{2011GeoRL..3802203B,2012GeoRL..3910203G}. 
In our model, we use altitude-dependent rates of production of hot oxygen, $f(E,h)$ by both DR of O$_2^+$ and precipitating ENAs, as described in the previous section. The attenuation of fluxes of hot O and OH is estimated based on their mean free path in thermal atmospheric gases. 
Such an approach is a compromise between a simple exobase approximation and more complex Monte Carlo simulation. In comparison, our multi-species 3D Monte Carlo simulation of the ENA-induced escape of hydrogen predicted about 40\% higher escape rates than a 1D model \citep{2014ApJ...790...98L}.

With the above assumptions, the volume production rate of hot OH$(v',j')$ at the altitude $h$ can be written as
\begin{eqnarray}
  P_{v' j'}(h) & = & \frac{1}{2} \int_{0}^{\infty} \mathrm{d}E \, T_{\mathrm{OH}}(h,E) 
      n_{\mathrm{H_2}}(h) \Gamma_{vj}^{v'j'}(E) \sigma_{vj,v'j'}^{\mathrm{mt}}(E) \nonumber  \\
             &   & \times \int_{h_{2}^{\mathrm{min}}}^{h} \mathrm{d}h_2 f(E,h_2) T_\mathrm{O}(h_2,h,E) ,
  \label{eq:P}
\end{eqnarray}
where $f(E,h)$ is the total hot O source function, $\Gamma_{vj}^{v'j'}(E)$ is the fraction of the OH molecules energetically allowed to escape, $\sigma_{vj,v'j'}^{\mathrm{mt}}(E)$ is the momentum transfer reactive cross section, and $n_{\mathrm{H_2}}(h)$ is the density of atmospheric H$_2$ gas taking part in reactions. The prefactor $1/2$ implies that one half of the produced energetic OH molecules are scattered towards the planet and cannot escape regardless of the energy transferred. 
The transparencies $T_{\mathrm{OH}}$ and $T_\mathrm{O}$ are defined as 
\begin{eqnarray}
  T_{\mathrm{OH}}(h,E) & = & \exp \left[-\int_{h}^{h_\mathrm{max}} \mathrm{d}h' \sum_i
                 \sigma_{\mathrm{OH},i}^{\mathrm{el,mt}}(E) n_{i}(h') \right] \nonumber \\
  T_\mathrm{O}(h_2,h,E) & = & \exp \left[ -\int_{h_2}^{h} \mathrm{d}h' \sum_i 
                      \sigma_{\mathrm{O},i}^{\mathrm{el,mt}}(E) n_{i}(h') \right]  ,
\end{eqnarray}
and describe the loss of flux due to the collisions with thermal atmospheric species. Specifically, $T_{\mathrm{OH}}$ is equal to the escape probability of hot OH$(v',j')$ produced in collisions with the incident hot O of energy $E$ at the altitude $h$, while the $T_\mathrm{O}$ is defined as the probability that the hot O atoms, produced at the altitude $h_2$, reach the altitude $h$ without the energy loss in collisions with other atmospheric constituents. 
The sums describe the loss of O and OH flux in collisions with major constituents of the martian upper atmosphere, where, for the $i$-th atmospheric species, $n_i(h)$ is the density, and $\sigma_{\mathrm{H_2},i}^{\mathrm{el,mt}}(E)$ and $\sigma_{\mathrm{O},i}^{\mathrm{el,mt}}(E)$ are elastic momentum transfer cross sections. Eight species were included in the summation: CO$_2$, CO, N$_2$, O$_2$, H$_2$, H, Ar, and H. We use only the elastic cross sections, which are about an order of magnitude larger than reactive or inelastic cross sections, to evaluate the transparencies. The integrals were evaluated using $h_{2}^{\mathrm{min}}$ and $h_{\mathrm{max}}$ as practical integration limits, taken to be 130 km and 800 km, respectively. 

We used the momentum transfer cross sections for OH$(v',j'$)+H scattering calculated in our previous work \citep{2014JChPh.141p4324G}. For collisions with the other species we approximated the cross sections by mass-scaling from known species: OH-He (from Ar-H$_2$), OH-N$_2$, OH-O$_2$, OH-CO (from O-H$_2$), and OH-CO$_2$ (from O-N$_2$ \citep{1998JGR...10323393B}).
Similarly, the cross sections for O-N$_2$, O-O$_2$, O-He, and O-H were adopted from \cite{1998JGR...10323393B}, \cite{1981JChPh..74.6734B}, \cite{2011GeoRL..3802203B}, and \cite{2009JGRA..114.7101Z}, respectively. As an additional simplification, we did not include O-O collisions in the flux attenuation since they do not reduce the total kinetic energy in the upward flux of hot O atoms. We also assume that all reactive collisions of O+H$_2$ take place with molecular hydrogen in its ground state, $(v=0,j=0)$, while at $T=280$ K, first three rotational states ($j=0,1,2$) will be significantly populated. We do not expect this approximation to introduce significant errors since the differences in reactive cross sections between these states are smaller than 3\%.

\section{Altitude profile of hot OH}

\begin{figure}[t]
  \noindent \includegraphics[width=0.95\columnwidth]{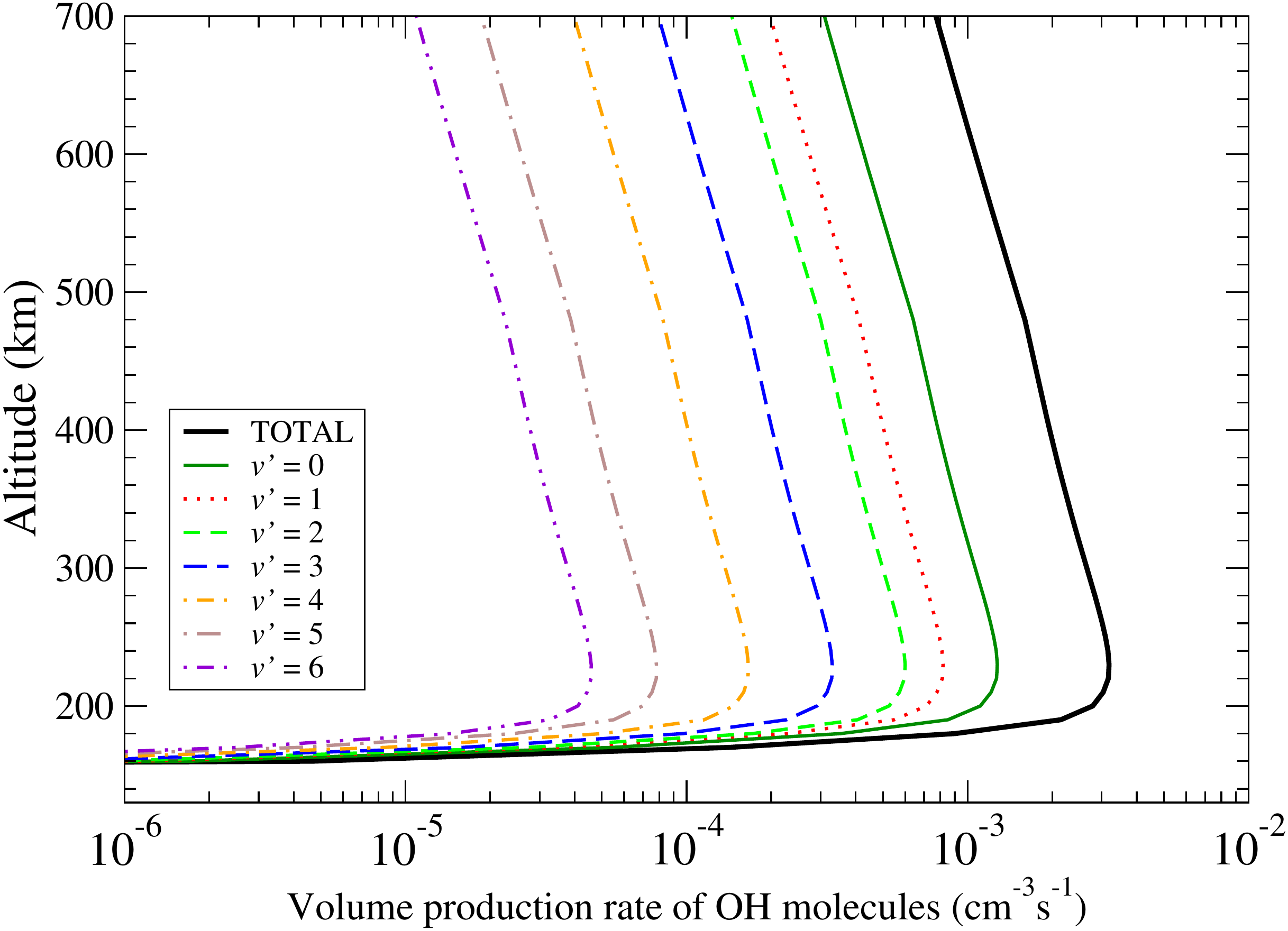}  
  \caption{Altitude profiles of volume production rates of vibrationally excited OH($v'=0 \dots 6$) produced in reactions with non-thermal O atoms. The rates are summed over rotational states $j'=0\dots36$ of OH($v',j'$).}
  \label{fig4a}
\end{figure}

\begin{figure}[t]
  \noindent \includegraphics[width=0.95\columnwidth]{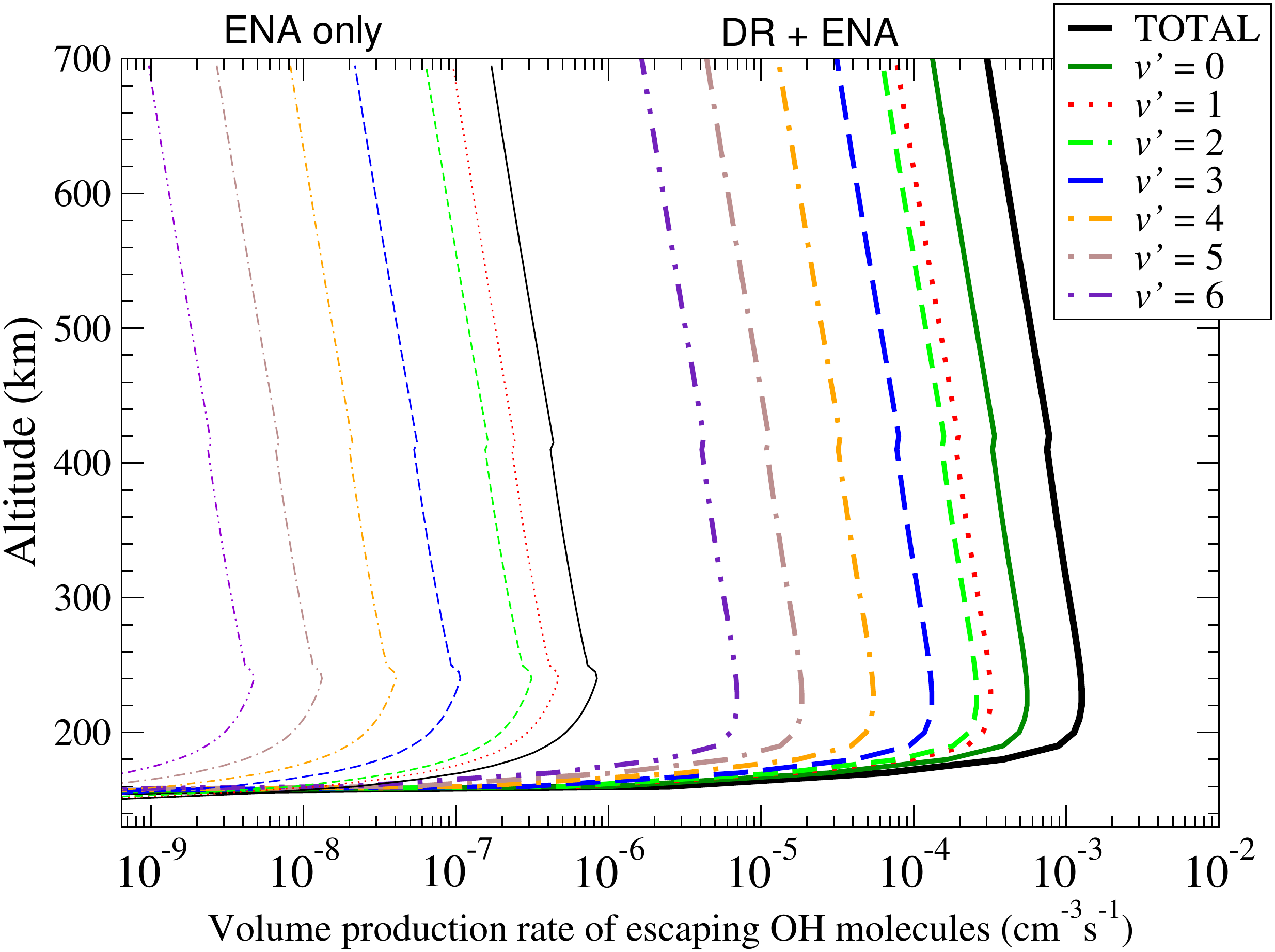}  
  \caption{Altitude profiles of volume production rates of vibrationally excited nascent OH($v'=0 \dots 6$) sufficiently energetic to escape produced in reactions with non-thermal O atoms. The rates are summed over rotational states $j'=0\dots36$ of OH($v',j'$) and shown for both sources of hot oxygen (thin lines for ENA).}
  \label{fig4b}
\end{figure}

Using Eq. (\ref{eq:P}) we have calculated volume production rate of hot OH$(v',j')$ molecules in the reaction H$_2(v=0,j=0)$ + O at the altitudes ranging from $h_{\mathrm{min}}=130$ km to $h_{\mathrm{max}}=700$ km, for all energetically-allowed target RV states, $v'=0 \dots 6$ and $j'=0 \dots 36$. We have included both sources of hot O, DR and collisions with H and He ions and atoms, and performed the calculation with and without (by setting the escaping fraction factor $\Gamma_{vj}^{v'j'}(E) = 1$) the constraint on the kinetic energy of the product OH.
The resulting altitude profiles of vibrationally excited hot OH, $P_{v'}(h) = \sum_{j'} P_{v'j'}(h) $, are shown in Figs. \ref{fig4a} and \ref{fig4b}, for the total and escaping volume production rate, respectively.

The volume production rates of hydroxyl molecules are very dependent on density of the atmosphere and availability of hot oxygen. They peak at about 210-220 km and remain significant below the exobase, up to about 170 km, below which the martian atmosphere becomes sufficiently dense to allow more efficient diffusive mixing and fast thermalization of hot oxygen. The production rates due to the ENA sources of hot O are about 500 times smaller than due to the DR, which is understood to be the dominant process of non-thermal escape. Our model predicts that the non-thermal reaction will mostly produce OH in $v'=0$ vibrational state, followed by higher vibrational states up to $v'=6$, which comprises less than 1\% of the population. A similar range of excited vibrational states is predicted for the OH energetically capable of escaping from Mars, even though the ratios between the individual excited states are somewhat different. This is particularly the case in enhanced rates for escaping OH$(v'=2)$, which are comparable to those for OH$(v'=1)$ state.

\begin{figure}[t]
\noindent \includegraphics[width=0.95\columnwidth]{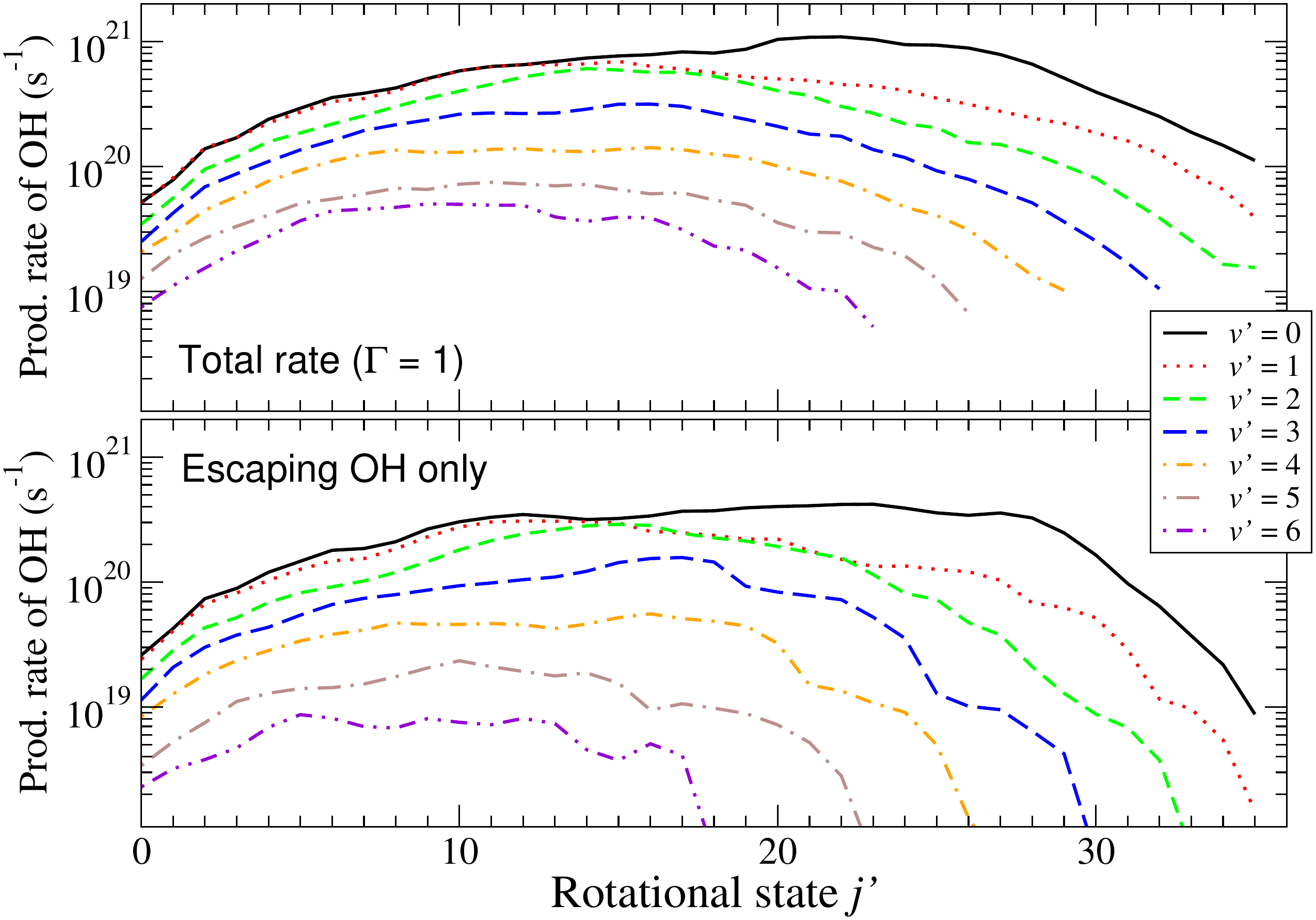}
\caption{Top: Total production rates of OH$(v',j')$ as a function of its rotational and vibrational states. Bottom: As above, for escaping OH($v',j'$). }
\label{fig5}
\end{figure}

We calculated the total production rates and the production rates of OH$(v',j')$ capable of escaping as the product of the surface area of the dayside of Mars and the non-thermal flux $\phi_{v'j'}$ given by:
\begin{equation}
  \phi_{v'j'} = \int_{h_{\mathrm{min}}}^{h_{\mathrm{max}}} P_{v'j'}(h) \mathrm{d}h 
\end{equation}
The total production rates of OH$(v',j')$ are shown in Fig. \ref{fig5}, where we also show distributions of rotational states $j'$ for each vibrational state $v'$. In case of the OH molecules sufficiently energetic to escape, the rates correspond to the number of OH molecules in a state $(v',j')$ potentially escaping from Mars at $h_{\mathrm{max}} = 700$ km and can be compared with other escape processes. On the other hand, the total production rate, unconstrained with respect to the kinetic energy, gives an indication of the total number of excited OH$(v',j')$ molecules distributed according to the altitude profiles given in Fig. \ref{fig4a} on the dayside of the planet and can be used to evaluate the brightness of the emission spectra and Meinel bands of OH \citep{1950ApJ...111..555M}.
The total escape rates of OH, summed over all internal states and compared with the rates for Jeans and non-thermal escape of O, H, and H$_2$ from literature, are given in Table \ref{table1}. 

\begin{table}
\centering
\begin{tabular}{r | c c  c  c}
\hline 
                                                                   & OH     & H$_2$    & O                  &   H     \\ \hline
   Jeans ($\times 10^{23}$ s$^{-1}$)          &           &  5.8$^a$  &                      & 530$^b$    \\
   NT ($\times 10^{23}$ s$^{-1}$) & 0.53   & 1.0$^a$  & 170-410$^c$ &           \\   
\hline 
\end{tabular}
\caption{Total non-thermal (NT) escape rates of OH from the ideal dayside of Mars calculated at 700 km, assuming Mars' volumetric mean radius equal to 3389.5 km and an ideal dayside. Published escape rates for H$_2$, O, and H are shown for comparison ($^a$\cite{2014JChPh.141p4324G}, $^b$\cite{2008SSRv..139..355J}, $^c$\cite{2015JGRE..120.1880L}).}
\label{table1}
\end{table}

The produced hydroxyl molecule will decay mostly by predissociation, followed by direct photo-destruction, while interactions with solar wind ions are not expected to play a major role. The predissociation rates vary up to about 30\% with solar activity, and give a range of OH lifetimes of (1.13-1.68)$\times 10^5$ sec for solar minimum conditions, considered in our calculation, and $(1.04-1.49) \times 10^5$ sec for the solar maximum, with uncertainties of up to 20\% \citep{1994Icar..107..164B}. The most important processes responsible for destruction of OH in the lower atmosphere of Mars include reactions with O, N, C, and H$_2$, with the reaction rates of the order of $(1-7)\times10^{-10}$ cm$^3$s$^{-1}$. The collisions with O will be a dominant mechanism of destruction at lower altitudes, where we can estimate the lifetime of OH to be of the order of 10-100 seconds, depending mainly on the atomic oxygen densities. This does not limit the escaping flux significantly.
The lifetime of excited OH in the $v'=2$ and $v'=1$ is about 12 and 24 ms \citep{1984Icar...59..305V}, respectively, which allows the OH formed in the upper atmosphere to travel between 60-75 km ($v=1$) or 115-150 km ($v=2$) before decaying, mainly by spontaneous emission. Higher vibrational states will decay faster, resulting in faint Meinel emission bands localized at the regions where the OH production rates are the largest.

\section{Conclusions}

We report the first theoretical model of non-thermal formation of excited OH molecules in the upper atmosphere of Mars in reactions of translationally hot O atoms and atmospheric H$_2$. The produced OH is very energetic, capable of escaping into space, and expected to contribute a small fraction of rotationally and vibrationally excited hydroxyl molecules to the Martian hot corona. These OH molecules are expected to have lifetimes up to several hours providing they are not destroyed in collisions. We estimate that the process contributes up to about 0.1\% of the total rate of escape of the Martian atmosphere to space.  
Our model is based on state-to-state energy dependent reactive cross sections for O+H$_2$ reaction and a 1D model of transport in the Martian atmosphere. The hot O atoms produced by photo-dissociative recombination of O$_2^+$ and collisions of thermal oxygen atoms with H and He ENAs were considered. The ENAs are found to contribute less than 1\% to the total formation rates for the present-day solar flux. 

This non-thermal process could be identified by its characteristic emission profile from high rotational and vibrational states of OH (vibrational states up to $v'=6$ could be excited). Recently, Meinel emission bands up to $(3-2)$ and $(3-1)$ were detected in limb observations of Mars' atmosphere with MRO CRISP instrument and taken as a signature of presence of OH in the Martian atmosphere \citep{2013Icar..226..272T}. The emissions are likely to originate from the altitudes between 45 and 55 km, where the concentration of the OH is the highest, and are described well by the existing models that include relevant chemical processes in middle atmosphere. While low densities of OH in the upper atmosphere present a significant difficulty for observations, identification of emissions from high Meinel OH bands, with the intensities following the predicted altitude profile, would confirm the presence of this non-thermal process. 
Moreover, the described non-thermal mechanism and detailed information about populated excited states may be helpful in interpreting high-resolution spectra of neutrals in the upper atmosphere of Mars, leading to more accurate estimates of total escape rates of H$_2$ and O from Mars.

We expect that the developed model can be adapted to other planetary atmospheres where the described high-temperature reactions can take place, including comets. 

\section{Acknowledgments}
M.G.'s research was supported by an appointment to the NASA Postdoctoral Program at the NASA Ames Research Center, administered by Universities Space Research Association under contract with NASA.





\bibliographystyle{elsarticle-harv} 
\bibliography{OH_mars}{}





\end{document}